# Grid-based Approaches for Distributed Data Mining Applications

**Lamine M. Aouad, Nhien-An Le-Khac, and Tahar Kechadi**

University College Dublin, School of Computer Science & Informatics
Belfield, Dublin 4 – Ireland
{lamine.aouad,an.le-khac,tahar.kechadi}@ucd.ie

**ABSTRACT**

The data mining field is an important source of large-scale applications and datasets which are getting more and more common. In this paper, we present grid-based approaches for two basic data mining applications, and a performance evaluation on an experimental grid environment that provides interesting monitoring capabilities and configuration tools. We propose a new distributed clustering approach and a distributed frequent itemsets generation well-adapted for grid environments. Performance evaluation is done using the Condor system and its workflow manager DAGMan. We also compare this performance analysis to a simple analytical model to evaluate the overheads related to the workflow engine and the underlying grid system. This will specifically show that realistic performance expectations are currently difficult to achieve on the grid.

## 1. INTRODUCTION

The grid has emerged as an important area in distributed and parallel computing. Increasing amounts of scientific communities are using grid facilities to share, manage and process large-scale datasets and applications. The grid can be seen as a computational and large-scale support, and even as a high-performance support in some cases. Furthermore, many high level grid-based knowledge discovery and data mining systems have recently been proposed [9][19][25][27]. These frameworks need efficient and well-adapted approaches able to generate an accurate global knowledge, based on distributed local mining and knowledge models, taking into account the high-end scalability of the grid.

Typical data mining applications include naturally distributed sources of data, as in customers' relationship management, marketing and sales, process modeling and quality control, genomic data and microarray analysis, text classification, among others. The underlying structures related to these applications are already grid-like and widely distributed. In this paper we present new approaches for two basic applications in data mining; namely clustering and frequent itemsets generation, for grid systems. These approaches are motivated by the inherent distributed nature of these applications and by the challenge of developing scalable solutions for the data mining field, taking into account the constraints related to analyzing massive distributed datasets on the grid. Performance evaluation is carried out on a configurable and controllable grid infrastructure which allows to monitor and reproduces a variety of experimental conditions.

Furthermore, grid-based data mining applications are likely to use federated and existing grid technologies which hide the complexity of multiple ownership, domains, and users. However, the algorithmic approaches used at higher levels are very important for scalability and optimization of the distributed processing cost. Well-adapted algorithmic approaches are then of prime importance in the design of data mining applications and frameworks for the grid. Indeed, we are already designing and implementing the ADMIRE framework which is a data mining engine on the grid intended to provide grid-based mining techniques and dynamicity at the user level. It also provides higher knowledge map representations of the mined data, both in local nodes and globally [27].

The rest of the paper is organized as follows, the next section surveys some distribution and grid-based efforts in the data mining area. Then, section 3 describes the proposed approaches. In section 4, we present the grid platform and the experimentation tools. Section 5 shows experimental evaluations, estimated results, discussions, and highlights directions for future work. Finally, section 6 concludes the paper.

## 2. BACKGROUND

Many research works addressed this area over the last two decades. Most of them were specially dedicated to parallel implementations of widely used algorithms and techniques on clusters of workstations or high performance computers using standard message passing interfaces such as MPI or PVM [2][8][11][13][15][17][18][24][31][33][34]. These implementations were mostly related to some clustering algorithms such as K-Means or its variants, or density based clustering algorithms (such as DBSCAN or BIRCH). The frequent itemsets generation task and association rules mining were also studied and a range of approaches were proposed including ODAM, FDM, the DDM approach, parallel FP-Growth, and other variants. However, most of the existing approaches in the literature need either multiple synchronization steps among processes or a global view of the dataset at some stages, or both.

Typically, distributed approaches act by producing local models followed by the generation of a global model by aggregating the local results. They are based on the global reduction, one or many times, of the so-called sufficient local models or statistics. Some works are presented in [20][21][22][23][32], mostly related to the clustering task or the association rules mining as quoted earlier. Furthermore, there are only few works intended to distributed systems using different computational sites and domains, with different architectures, dynamic resources, etc. We can cite, for frequent itemsets generation, the GridDDM approach [35] which mines maximal itemsets from distributed datasets. Other recent works for widely distributed data mining algorithms include association rules mining [39], or the K-Means clustering algorithm [40]. There exists several papers summarizing the work that has been done in distributed data mining such as in [36], or in [37] and [38] dedicated to association rules mining.

Furthermore, many grid-based projects and frameworks already exist or are being proposed in this area such as Knowledge Grid [9], Discovery Net [19], Grid Miner [25], ADMIRE [27], etc. These tools aim to offer a high level abstractions, and techniques for distributed data mining and knowledge extraction from data repositories or warehouses available on the grid. It generally uses basic grid mechanisms, mainly provided by existing grid environments, to build their specific knowledge discovery services. Many of them use services provided by the Globus Toolkit. Furthermore, beyond their architecture design, the data analysis and management policies, the integration or placement approaches, the underlying middleware and tools, etc. the grid needs efficient and well-adapted algorithms. The motivation of this work is to propose and test two new grid-based techniques in real distributed multi-sites conditions.

## 3. GRID-BASED DATA MINING ALGORITHMS

In this section we describe the proposed approaches; namely a variance-based distributed clustering and a grid-based frequent itemsets generation.

## 3.1 Distributed clustering

Clustering is a fundamental task in the data mining field. It groups data objects based on information found in the data that describes the objects and their relationships. The goal is to optimize similarity within a cluster and the dissimilarities between clusters in order to identify interesting structures in the underlying data. There is already a large amount of literature in the field ranging from models, algorithms, validity and performances studies, etc. Details about these works are out of the purpose of this paper, we refer the reader to [41] and [42] where detailed overviews can be found.

The key idea of the proposed approach is to perform local clustering with a relatively high number of clusters, which are referred to as sub-clusters, or an optimal number of clusters found by using an approximation technique (such as the Gap Statistic technique), then to merge them at the global level according to an increasing variance criterion which requires a very limited communication overhead. In this algorithm, all local clustering are independent from each other and the global aggregation can be done independently, at any initial local process. The merging process of local sub-clusters at the global level exploits then the locality in the feature space, i.e. the most promising candidates to form a global cluster are sub-clusters that are the closest in the feature space, including sub-clusters from the same node.

An important aspect of this algorithm is that the merging is 'logical', i.e. each local process can generate a global labeling among local sub-clusters, without necessarily reconstructing the overall clustering output, which means without communicating sub-clusters. That is because the only bookkeeping needed from the other sites is centers, sizes and variances. The aggregation is then defined as a labeling process among local sub-clusters in each participating site. This can be followed by a broadcast of the results if needed. Furthermore, a perturbation process is performed when the merging action is no longer applied. A number of candidates (user defined parameter) are collected from each global cluster at its border (proportional to the overall size composition). Then, the process moves these candidates by trying the closest ones and with respect to the gain in the variance criterion when moving them from the neighboring global clusters. The algorithm is described bellow.

**Algorithm 1.** Distributed clustering.
**Input:** $X_i$ ($i = 1,...,s$) datasets, and $k_i$ the number of sub-clusters in each site $S_i$
**Output:** $k_g$ global clusters, i.e. the global labeling

```
 1: for i = 1 to s do
 2:     LS_i = cluster(X_i,k_i)
 3: end for
 4: j = select_aggr_site()
 5: for i = 1 to s do
 6:     if i ≠ j then
 7:         send(,,,j)
 8:     end if
 9: end for
    At site j:
10: while var(C_i,C_j) < ∫_{ij}^{max} do
11:     merge(C_i,C_j)
12: end while
13: for i = 1 to k_j do
14:     find_border(b,i)
15:     add_multi_attributed(i)
16:     for x = 1 to b do
17:         j = closer_global(x)
18:         var_new = var(C_i - C_x, C_j + C_x)
19:         if var_new < var then
20:             label(x, j)
21:             var = var_new
22:         end if
23:     end for
24: end for
```

In the first step, local clusterings are performed independently on each local dataset. The local number of clusters can be different in each site. Each local clustering in a given site $i$ gives as output $k_i$ sub-clusters identified by a unique identifier $cluster_{i,number}$ with $number = 1,..,k_i$, and their sizes, centers and variances. At the end of local processes, local statistics are sent (lines 5 to 9) to the merging process $j$ (4). Then, sub-clusters aggregation is done in two phases; merging (10 - 12) and perturbation (13 - 24). In the latter phase, the border $B_i,b$ is found (14 - 15), with $i = 1,..,k_g$, and $b$ is a user defined parameter. For each $x$ in $B_i,b$, the closet global cluster $j$ is found and the new variance is computed. The actual perturbation is done if the new global variance is smaller than the initial one (16 - 23) [7]. Recall that this process is a labeling of local sub-clusters at the global level. At the line 11, the new global statistics, namely the size, center and variance, are:

$$N_{new} = N_i + N_j$$

$$c_{new} = \frac{N_i}{N_{new}} c_i + \frac{N_j}{N_{new}} c_j$$

$$var_{new} = var_i + var_j + \mathbf{s}(i,j), \quad \forall C_i, C_j, i \uparrow j$$

where

$$\mathbf{s}(i,j) = \frac{(N_i N_j)}{(N_i + N_j)} \cdot d(c_i, c_j)$$

represents the increase in the variance while merging $C_i$ and $C_j$, and $d$ represents the Euclidean distance.

### 3.2 Distributed frequent itemsets mining

Discovering frequent itemsets is a crucial task in data mining. It is a core step of various tasks including association rules, correlations, causality, and episodes, among others. Many existing algorithms, both sequential and parallel, are related to the Apriori algorithm [3]. This algorithm exploits the observation that all subsets of a frequent itemset must be frequent. We propose here a grid-based implementation based on this algorithm.

In the proposed approach, the local generation of the candidate sets is in the Apriori manner. The difference from the existing approaches is that frequent itemsets of the requested size are computed locally without any global pruning strategy, and the global frequency is deduced at the end of local processes using a single global phase. Indeed, a grid-based implementation implies some constraints related to the underlying middelware and tools, and communication and synchronization overheads, which are excessive and consequently highly suitable to avoid. This approach greatly reduces the communication costs and avoids multiple synchronization phases (reduced to only one). Algorithm 2 describes this approach.

In the first step, every site performs an Apriori generation of sizeindependently of each other. This means that only local pruning is considered to generate locally frequent itemsets. A single synchronization step is required at the end of this step to generate the global frequency. Remote support counts of locally frequent itemsets and subsets of those that fail the global frequency test are requested from other sites iteratively. All sizes are then generated by a top-down search. This requires only a few communication passes.

Our approach is motivated by the performance behavior of the Apriori generation process [6] which shows that communication steps and computations of remote support counts are computationally inefficient, and constrains the global performance. This is due to the fact that this generation task is level-dependent, which means described by a switch level from constantly high success rates to very low success rates, and usually a switch to zero. Basically, this means that intermediate global pruning strategies do not bring enough useful information locally. A comparison with a classical Apriori-based algorithm, namely the FDM approach (Fast Distributed Mining of association rules) [11] is given in order to argue our approach for the grid.

**Algorithm 2.** Grid-based Frequent itemsets Mining (GFM).
**Input:** $X_i$ ($i = 1,..., s$) datasets and $k$
**Output:** Globally frequent itemsets of sizes 1 to $k$

1: **for** $i = 1$ to $s$ **do**
2:     apriori_gen($X_i$, $k$)
3: **end for**
4: **for** $i = 1$ to $s$ **do**
5:     **while** $LL_i \neq \emptyset$ **do**
6:         **for** $j = 1$ to $s$, ($i \neq j$) **do**
7:             receive_remote($i$, $LL_i$)
8:             local_support($LL_i$)
9:             send_support($i$, $j$)
10:        **end for**
11:        **for** j = $1$ to $s$, ($i \neq j$) **do**
12:            receive_remote_support($j$)
13:        **end for**
14:        $L_i$ += generate_globally_frequent()
15:        $LL_i$ = subsets_globally_failed()
16:    **end while**
17: **end for**
18: $L = U_{i=1}^{s} L_i$

The main goal of this algorithm is to avoid many synchronizations and communications steps, required in most of the existing distributed approaches, rather than minimizing local execution times or considering constraints in the mining process. Indeed, these synchronization steps can generate very costly overheads in the grid. Furthermore, we will show that the top-down approach generate less communication needs than the classical down-up approach used in classical implementations. Also, it is well known that the candidate sets can drastically grow in the Apriori-like algorithms leading to memory constraints. This means that it is usually not possible to ship all existing data to one site since the data could not fit in memory. This natural candidate for distribution needs then an efficient grid-based approach.

## 4. EXPERIMENTATION TOOLS

In this section we present the Grid'5000 platform and the grid framework used to manage the computational resources.

### 4.1 The Grid'5000 platform

Grid'5000 [4] [29] is a large scale environment for grid research. It aims to provide a reconfiguration and monitoring instrument to investigate grid issues under real and controllable conditions. Grid'5000 is not a production grid; it is an instrument that can be configured to work as a real testbed at a wide area scale. Grid'5000 is composed of nine geographically distributed clusters. A set of control and monitoring tools allows users to make reservation, configure or reconfigure a specific owner environment, make deployment, and run experiment and measurements. It is already used in many experiments at different levels of the grid including network protocols, operating system mechanisms, middleware, issues in performance, scheduling, fault tolerance and parallel and distributed programming and applications.

### 4.2 Workflow management in the grid

Several significant research works have been conducted in recent years to automate the workflow activities using advanced workflow management tools in the grid. The concept of workflow or process arrangement is extremely important and useful for a large range of applications within the grid context, including basic data mining tasks. A large number of tools, with a large panel of capabilities, have been proposed and used by the community, including the Condor DAGMan meta-scheduler [30], YvetteML [12], Askalon [16], GridAnt [5], among many others. A number of description languages have been proposed to declare the jobs composition, many of them have an XML based modeling. The architecture is usually composed of a user interface or language tools and the workflow execution engine which controls the execution within the grid. In the next paragraph, we briefly present the Condor system and its DAGMan workflow manager (Directed Acyclic Graph Manager).

### 4.2.1 Condor/DAGMan

Condor is a batch system providing a job management mechanism, resource monitoring and management, some scheduling capabilities and priority schemes [30]. The Condor system provides a ClassAds mechanism for matching resources requests and offers, checkpointing and migration mechanisms, and job management capabilities to across the grid with Condor-G (using the Globus Toolkit), and Condor-C which allows jobs to be moved between machines job queues.

The DAGMan layer provides a directed acyclic graph representation manager which allows to express dependencies between Condor jobs. It allows a user to list the jobs to be done with constraints on the order through several description files for the DAG and the jobs within the task graph. It also provides fault-tolerant capabilities allowing to resume a workflow from where it is left off, in the case of a crash or other failures. However, the scripting language required by DAGMan is not flexible since every job in the DAG has to have its own condor submit description file.

## 5. EXPERIMENTAL EVALUATIONS AND MODELS

### 5.1 Experimentation setup

The computational resources, used in our tests, are distributed over five sites of the Grid'5000 platform. A brief description of these nodes is given in Table 1. Local clusters are interconnected via Gigabit Ethernet and the different locations by RENATER through VLANs using MPLS at level 2. The average values of bandwidth and latency between the computing sites, measured using the NetPerf tool [1], range between 12.71 Mb/s and 106.63 Mb/s for bandwidth, and between 5 and 28 ms for latency. In local sites the averages bandwidth is about 941 Mb/s and 0.07 ms for latency. Table 2 gives the average observed during the tests and used in the estimation process below.

Table 1. The grid testbed.

| # used per site | CPU/Memory | Location |
|---|---|---|
| 200 | AMD-64 opteron, 2 GHz/2GB | Orsay |
| 70 | bi-proc AMD-64 opteron, 2 GHz/2GB | Sophia |
| 64 | bi-proc AMD-64 opteron 248, 2.2GHz/2GB | Rennes |
| 58 | bi-proc AMD-64 opteron 248, 2.2GHz/2GB | Toulouse |
| 46 | bi-proc AMD-64 opteron 246, 2.2GHz/2GB | Nancy |

Table 2. Average bandwidths and latencies among the different sites involved during tests.

| Mb/s - ms | Orsay | Toulouse | Rennes | Nancy | Sophia |
|---|---|---|---|---|---|
| Orsay | - | 16.15 - 15 | 57.73 - 8 | 90.77 - 5 | 17.63 - 28 |
| Toulouse | 38.97 - 15 | - | 26.08 - 19 | 28.89 - 17 | 35.74 - 14 |
| Rennes | 66.33 - 8 | 12.71 - 19 | - | 44.63 - 11 | 26.96 - 19 |
| Nancy | 106.63 - 5 | 14.13 - 17 | 44.54 - 11 | - | 30.01 - 17 |
| Sophia | 21.45 - 28 | 17.41 - 14 | 26.93 - 19 | 30.14 – 17 | - |

The configuration of the computational nodes is made up in 3 steps starting by the reservation of the required nodes in each cluster. Initially, the required software, namely the condor system, is installed in a deployed node. Then, the environment is recorded for next deployments. The same process is done on each site since the computational recorded environments are not shared currently. The last step is the deployment of these environments in each site and their configuration through a set of shell scripts.

**5.2 Tests**

The tests are divided into two parts: experimental and analytical. The reason is that an important overhead related to the workflow engine and the underlying environment has been noticed in the first set of tests. This will be highlighted in the rest of the paper by giving an estimation of these overheads. The analytical part is then intended to give boundary results and an evaluation of the proposed algorithms without connecting them to the grid middleware performance. For both applications, synthetic datasets are generated. For the clustering task, the data is a set of random Gaussian distributions. For the frequent itemsets mining, synthetic transactions from different sizes were generated. Descriptions and sizes are given in the next section.

### 5.2.1 Experimental results

In this section, we will give execution times and details about each test. Overheads estimation related to the middleware, communications and the workflow engine will also be given and discussed later on.

For the frequent itemsets generation task, the tested dataset is composed of $4 \cdot 10^6$ transactions distributed over up to 200 processes (with about 20000 transactions each). For comparison purpose, the FDM approach was implemented. This algorithm uses interesting relationships between locally frequent and globally frequent itemsets, to generate smaller sets of candidates at each of the $k$ communication steps [11]. In these tests, the globally frequent itemsets of size 1 to 4 were generated. The average computing time for the proposed version was 521 minutes, and 687 minutes for the FDM approach. This corresponds to up to 25% better computing time for the proposed approach. Note that only 2 communication passes (instead of 4) were required for the proposed technique.

We also noticed that, in addition of the multiple communication and synchronization steps required in the FDM version, the remote support computation is quite computationally expensive, representing up to 13% of its whole computing time. Since this process is repeated $k$ times, this will decrease the performance compared to the proposed version especially for small sizes when the success rate is high. Also, the gain factor will be more important when $k$ increases.

For the clustering task, the dataset is composed of $5 \cdot 10^7$ samples distributed over 200 processes. Initial local clusterings use the *K-Means* algorithm with 20 sub-clusters in each process. For the merging step, the constraint parameter is set up as twice the highest individual sub-clusters' variance. This step includes sub-clusters perturbation once the merging is no longer applied, but no additional communications are required for the computation of the new statistics at this level.

The average execution time for the global clustering is 1050 seconds, including the initial communications for the input data files and all the job preparation and scheduling steps. The actual computing time for the clustering

(the maximum computed locally) and the merging processes represents approximately only 2% of the whole workflow execution time. We will discuss this issue further in the next section by giving analytical estimations.

5.2.2 Analytical estimations

In this section, we consider previous measurements and results to estimate the overheads. Indeed, we take into account the sizes of the candidate sets exchanged during support counts collections and job preparation and scheduling steps. For the frequent itemsets generation, the simple analytical model presented here considers execution times of the basic tasks needed in each process at each level. This includes the Apriori generation, support counts computation and the global frequency generation. Recall that the estimation of the communication times uses measurements of bandwidths and latencies obtained by the NetPerf network benchmark [1]. For the proposed version, if the application is intended to be executed in $p$ processes, the stages corresponding to the Apriori generation, support counts computation and local globally frequent itemsets generation, are then parallel activities. The overall time execution is the sum of the maximum execution times at each stage.

In the FDM implementation, all frequent itemsets are incrementally generated until the requested size (or the maximal one). If it is intended to be executed on $p$ processes and $k$ is the requested size, then there are $2k+1$ stages of parallel activities. As before, the overall estimated time is the sum of the maximum execution times at each stage (involving different processes, parameters, and inputs). Recall that the estimation of the communication times is based on the size of the real remote support counts requests files sent in the experiment tests described in the previous paragraphs.

Table 3. Results summary.

| Task | Calculated times | Estimated times | Estimated overhead |
|---|---|---|---|
| V-Clustering | 1050 | 19.52 | 98% |
| GFM | 521 | 424 | 18.6% |
| FDM | 687 | 518 | 24.6% |

The estimated times for both versions are respectively 424 and 518 minutes for $k$=4, with same sizes and local support bound as in the experimentation tests, since the performance of the Apriori generation, remote support counts computation and all intermediate processes can vary greatly depending on these parameters. The proposed version gives a better estimated time by 18.2% compared to the FDM approach. Recall that the global frequency is generated at the end of the overall local generation processes in the proposed version. Besides, this corresponds to an overhead between 18% and 25% respectively compared to the execution on the grid.

In the clustering case, the estimated execution time is the sum of the maximum of the execution times of local clustering processes, and the merging process. This time is about 19 seconds. Furthermore, according to Table 2, the worst communication overhead case gives an estimation of 0.52 second. The sum represents less than 2% of the whole execution time presented in the previous section. Also, the communication overhead for the aggregation step of our algorithm is very small compared to the computation times since the only statistics about local clustering transmitted to the aggregation process are centers, sizes and variances. Thus, more than 98% of the whole computing time is considered as overheads at different levels of the grid implementation. Results summary is given in Table 3.

### 5.3 Discussion

Our implemented algorithms, within a workflow environment for the grid, considerably reduce data communication and task synchronization which can be the most critical in term of execution efficiency since we consider these proprieties as the most suitable for grid-based applications. Still, some constraints on the experimental grid middleware make realistic expectations difficult to achieve and lead to relatively poor performance and scalability even with large and independent parallel loops. The job preparation and submission latencies are of prime importance in this case and are the first sources of efficiency loss. Indeed, even in a local execution test (in a laptop, Genuine Intel Centrino Duo 2GHz and 2GB memory) and with a small workflow application containing only two small jobs under Condor and DAGMan, the preparation process takes in average about 295 seconds (about 5 minutes), which represents the interval between the workflow launching, and the first job submission. However, this does not seem to be related to the size of the DAG, and can be variable as well. However, this gives an idea about the need to speedup this process in workflow environments.

Nevertheless, the comparison between our approach and the FDM algorithm for frequent itemsets generation and the estimated results show the effectiveness of the presented algorithm, and give good targets for future evaluations. As for execution overheads, we can notice the important gap between the execution times and the bounds given by estimated values, which can be seen as ideal execution times. The severity of the clustering case may comes from the fact that the local clustering is not computationally expensive, and thus the parallel execution section containing these tasks cannot improve the overall execution time. Grid tools are not optimized for such use.

The frequent itemsets generation task gives less important overheads (although up to 25%). This seems to show a great correlation between the size of the jobs in the workflow and its execution performance independently of the size of any parallel loop within the workflow. Some research works already tackled such performance issues for an efficient use of the grid at higher levels, especially for workflow programming. In [14] for instance, authors propose a decentralized architecture that aims to simplify and to reduce the overhead for managing large workflows through partitioning, improved data locality, and reduced workflow-level checkpointing overhead. In [28] authors give a detailed overhead analysis and formalism under the Askalon environment [16] which intends to understand the sources of bottlenecks involved in the loss of performance for workflow applications in the grid. Other works, especially in the business field, tackle the quality of service management of workflow and investigate the performance of various workflow architectures [10][26].

It is hard to estimate an ideal performance time for a workflow application on the grid since a hierarchy of overheads is involved and traditional parallel metrics are no longer applied. While it is mostly a complex scheduling issues for complex workflow applications, it seems to be more related to the grid tool structure and implementation in this case since the jobs were equivalent, almost in parallel loops which did not exceed the number of processors available. In addition, this required a very limited communication and synchronization needs, and was done under deployed environments, i.e. controlled and stable conditions. Our aim is not to explain the nature and reasons of such performance loss but to emphasize this open issue that still receives small consideration from the community through the implementation of real-world grid intended applications. However, this discussion is based on the Condor system and its DAGMan meta-scheduler. Indeed, other performance analysis have to be done under different grid middleware and workflow engines, even most of them are known to generate important overheads. Furthermore, analyzing the overall overhead as a succession or hierarchy of overhead could give more details about the efforts that developers and algorithms designers should focus on.

## 6. CONCLUSION

In this paper, we presented grid-based implementations of basic data mining applications; namely clustering and frequent itemsets generation. While an important effort has been made in sequential and parallel data mining applications, there are only few algorithms which tackle the distribution problem in loosely coupled environments. We evoked the need of efficient grid-based algorithmic approaches. We proposed lightweight distributed algorithms, for the clustering task, based on an increasing variance constraint, and a frequent itemsets generation technique which reduces the number of synchronization steps needed. Both of the proposed approaches have very limited communication overheads. Indeed, it is generally considered that an efficient grid-based algorithm needs to avoid multiple synchronization and communication phases as much as possible. Our global knowledge generation approaches follow such a scheme.

The results of the experiments and the comparison show good performance and substantial improvements in the implementation efficiency of the tested applications. Grid implementations have an essential need to exchange few data and avoid synchronization steps. However, realistic performance hopes still difficult to achieve in the grid since the gap between the estimated performance and experiments reaches 98% of the execution times (for the clustering case, excluding the communication overhead), even with very large independent parallel loops. This gap is less severe with more computationally expensive jobs (for the frequent itemsets generation). This is due to the fact that one of the most costly overhead, namely job preparation and scheduling, is partly overlapped by computations in the DAG. More analysis of these overheads would give more details about the efforts that should be done at different levels of the grid.

## ACKNOWLEDGMENT

*Experiments presented in this paper were carried out using the Grid'5000 experimental testbed, an initiative from the French Ministry of Research through the ACI GRID incentive action, INRIA, CNRS and RENATER and other contributing partners, which the authors would like to thank.*